\documentclass[11pt]{article}
\usepackage{graphics}

\begin{document}

\title{Counterexamples of Boltzmann's equation}
\author{C. Y. Chen\\
Department of Physics, Beijing University of Aeronautics\\
and Astronautics, Beijing, 100083, P.R.China}
\thanks{Email: cychen@buaa.edu.cn}
\maketitle

\vskip 25pt \centerline{PACS 51.10.+y. Kinetic and transport
theory of gases} \vskip 5pt

\begin{abstract}
To test kinetic theories, simple and practical setups are
proposed. It turns out that these setups cannot be treated by
Boltzmann's equation. An alternative method, called the
path-integral approach, is then employed and a number of
ready-for-verification results are obtained.
\end{abstract}

\section{Introduction}One of the fundamental mysteries in statistical
physics is that the derivation of Boltzmann's equation explicitly
resorts to the time reversibility of Newtonian mechanics whereas
the derived equation seems to offer a plausible account for time
irreversibility. This logic puzzle has been around us for more
than a hundred years and a large number of interpretations
motivated to clear up the issue kept on appearing in the
literature.

Without much attention received, we took another direction and
questioned the mathematical and physical validity of Boltzmann's
equation\cite{chen1}. In this paper, we concern ourselves with
proposing simple and practical setups, in which Boltzmann's
equation, or any other alternatives, can be put to computational
and/or experimental tests.

Two types of collisions get involved: particle-to-boundary
collisions and particle-to-particle collisions. In view of that
every particle can be thought of as a particle scattered by
boundaries or by other particles, the generality of this
discussion is rather obvious.

It turns out that for the proposed setups Boltzmann's equation
yields either incomputable or unreasonable results. With that in
mind, an alternative method, called the path-integral approach, is
employed and a number of ready-for-verification results are
derived.

If interested enough, readers may devise their own ways to prove
or disprove every conclusion offered by this paper.

\section{Two setups}

To test Boltzmann's theory, as well as its alternatives, we
advance the following two setups, which can be easily realized in
real or computer-simulated experiments.

 \begin{itemize}
\item {\bf Setup 1}: Referring to Fig.~1, consider a parallel beam
 whose distribution is
\begin{equation}\label{beam1} f'({\bf v}')=n'\cdot
g(v'_{x})\,\delta(v'_{y}-0)\,\delta(v'_{z}-0), \end{equation}
where $v'_{x}>0$ and $g(v'_{x})$ is a nonnegative normalized
function; and let it hit a solid boundary of finite size. It is
obvious that the distribution given above is quite `normal'; for
instance the relationship
\begin{equation} \int f'({\bf v}') d{\bf v}'\equiv
\int f'({\bf v}') \, dv'_{x} dv'_{y} dv'_{z} =n'
\end{equation} exists, where $n'$ is the ordinary particle density of
the beam.

\includegraphics*[150,585][500,725]{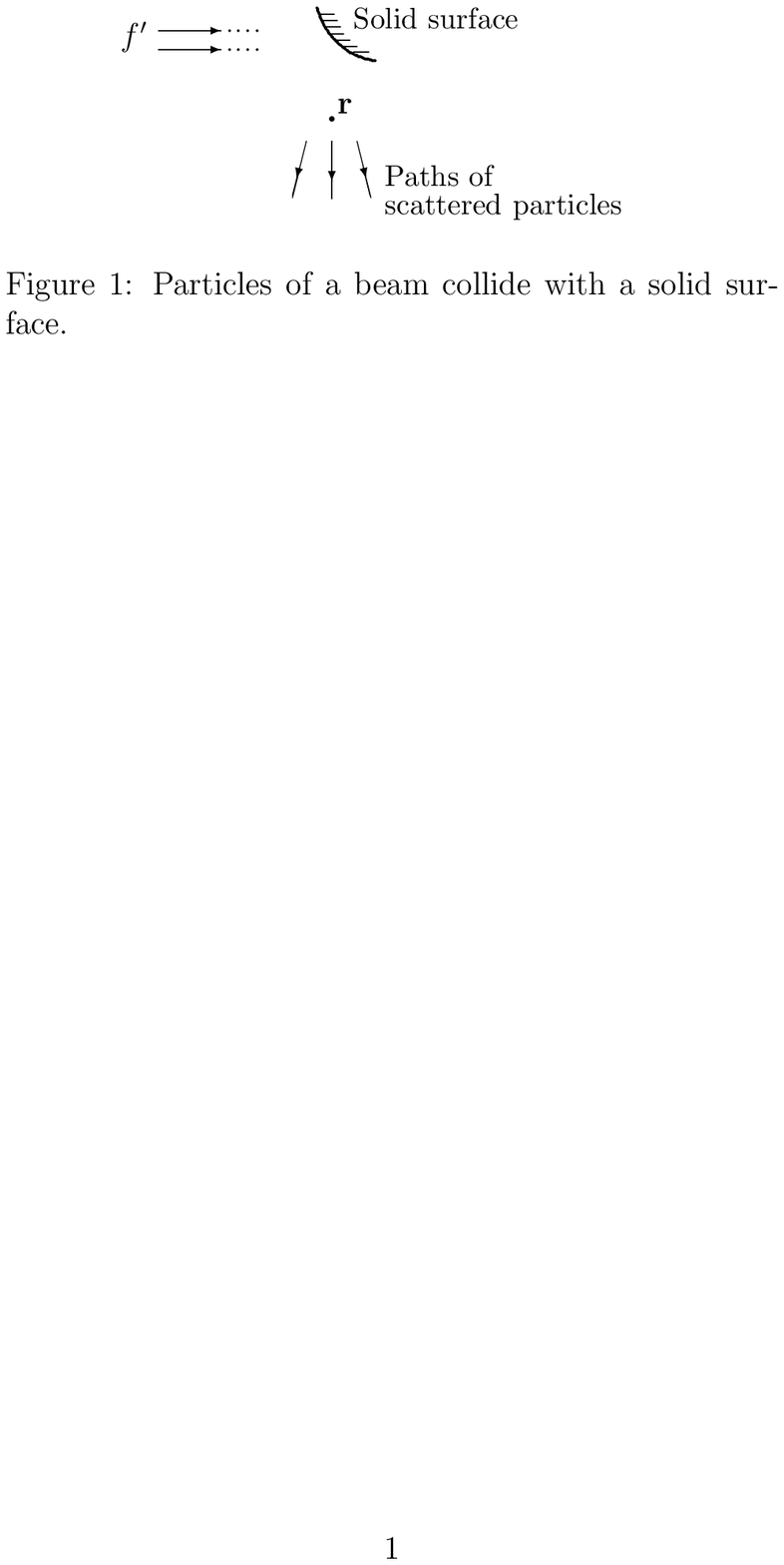}

 \item
{\bf Setup 2}: Consider two head-on parallel beams shown in
Fig.~2. Beam 0 which moves rightwards is described also by the
distribution function (\ref{beam1}); and beam 1 is described by
\begin{equation}\label{beam2}f'_1({\bf v}'_1)=n'_1\cdot
g_1(v'_{1x})\,\delta(v'_{1y}-0)\,\delta(v'_{1z}-0),
\end{equation} where $v'_{1x}<0$ and $g_1(v'_{1x})$ is again a
nonnegative normalized function. Though having the same mass, the
particles belonging to beam 0 and beam 1 (sometimes referred to as
type 0 and type 1 particles respectively) are considered to be
distinguishable in this paper. The transverse sections of the two
beams are finite, say, shaped like circular ones.

\includegraphics*[150,580][500,720]{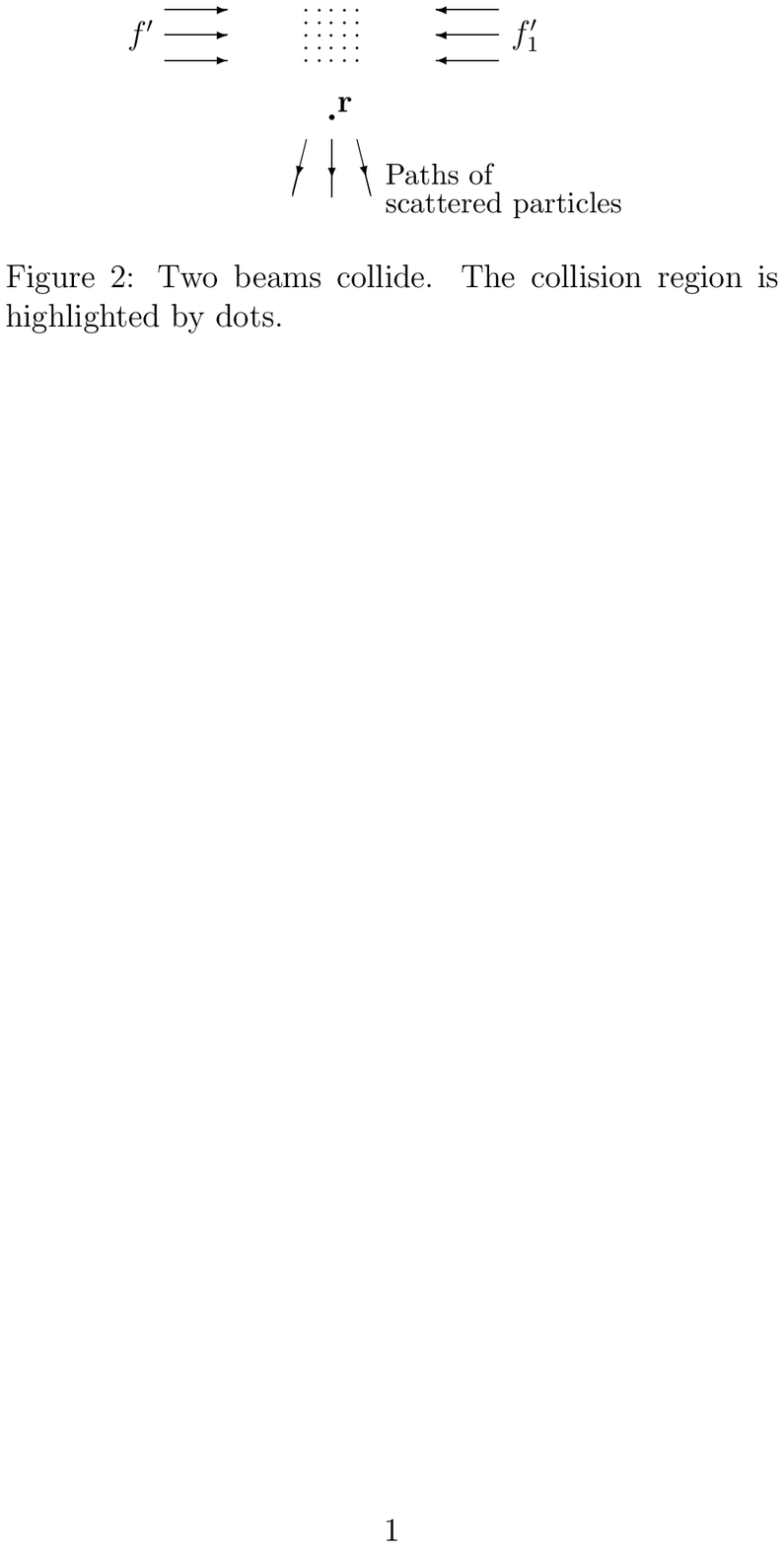}

\end{itemize}

To make the discussion less complicated, we shall ignore any
external forces and disregard particles that suffer collisions
twice or more (letting $\sigma$ be small and/or the gas be
dilute). With the assumption that all colliding parallel beams are
produced by stable external sources, the two setups are considered
to be time-independent.

\section{Application of Boltzmann's equation}

According to the standard theory\cite{reif,kubo}, in a spatial
region where collisions are ignorable, the distribution function,
denoted by $f(t,{\bf r},{\bf v} )$ with $\bf r$ representing the
position ($x,y,z$), satisfies the collisionless Boltzmann equation
\begin{equation}\label{bol0} \frac{\partial f}{\partial t}+
{\bf v}\cdot\frac{\partial f}{\partial {\bf r}}+ \frac{{\bf
F}}m\cdot \frac{\partial f} {\partial {\bf v}}=0,
\end{equation}
in which $\bf F$ stands for the external force acting on the
particles. In a region where collisions cannot be ignored, the
regular Boltzmann equation reads
\begin{equation}\label{bol} \frac{\partial f}{\partial t}+
{\bf v}\cdot\frac{\partial f}{\partial {\bf r}}+ \frac{{\bf
F}}m\cdot \frac{\partial f} {\partial {\bf v}}=\int_{{\bf
v}_1}\int_{\Omega_c} 2u (f'f'_1-ff_1) \sigma(\Omega_c) d\Omega_c
d{\bf v}_1, \end{equation}  where $f'({\bf v}')$ and $f'_1({\bf
v}'_1)$ represent the particles that collide to make $f({\bf v})$
increase, $f_1({\bf v}_1)$ represents the particles that collide
with $f({\bf v})$ to make $f({\bf v})$ decrease, $(2u)$ is the
relative speed of the colliding particles , $\Omega_c$ is the
solid angle formed by the initial relative velocity and final
relative velocity, and $\sigma(\Omega_c)$ is the cross section in
terms of $\Omega_c$.

In the following two subsections, we shall apply (\ref{bol0}) and
(\ref{bol}) to the setups given in Section 2.

\subsection{Application to setup 1}
It is trivial to see how the parallel beam illustrated in Fig.~1
obeys the collisionless Boltzmann equation (\ref{bol0}). Noting
that the beam is stationary and there is no external force, namely
$ {\partial f'}/{\partial t}=0$ and ${\bf F}=0$, we obtain from
(\ref{bol0})
\begin{equation} \label{solution01}
{\bf v}'\cdot \frac{\partial f'({\bf v}')}{\partial {\bf r}}=0
\quad{\rm or}\quad \left. \frac{\partial f'}{\partial {\bf
r}}\right|_{\rm path}=0,
\end{equation} which is indeed satisfied by (\ref{beam1}). For later
use, it should be mentioned that (\ref{solution01}) characterizes
parallel beams.

We now turn our attention to the particles scattered by the solid
surface, whose distribution function will be denoted by $f(t,{\bf
r},{\bf v})$. Again, due to the existence of $\partial f/\partial
t =0$ and ${\bf F}=0$, the collisionless Boltzmann equation
(\ref{bol0}) yields
\begin{equation} \label{solution11}\left. \frac{\partial
f}{\partial {\bf r}}\right|_{\rm path}=0\quad{\rm or}\quad f|_{\rm
path}={\rm Constant}. \end{equation} The deduction of
(\ref{solution11}) is simple, but difficult things arise. As
Fig.~1 has intuitively shown, the scattered particles will not
form a parallel beam and their density will in general decrease
along the paths (which can be easily verified in experiments), but
according to (\ref{solution11}) they seem to form a parallel beam
and their density seems to keep constant along any path of the
scattered particles. If we wish to adopt (\ref{solution11}), for
whatever reason, we shall encounter great difficulty in
interpreting it, let alone verify it in experiments.

\subsection{Application to setup 2}

In this subsection, $f(t,{\bf r},{\bf v})$ denotes the
distribution function of scattered beam 0 particles shown in
Fig.~2.

Firstly, we examine the situation outside the collision region.
Again, the collisionless Boltzmann equation gives rise to
\begin{equation}\label{solution1} {\bf v}\cdot
\frac{\partial f}{\partial {\bf r}}=0\quad{\rm or}\quad f|_{\rm
path}={\rm Constant}. \end{equation} Some of the problems related
to (\ref{solution1}) have been discussed, of which one is that the
density of the scattered particles will decrease along the paths,
 while (\ref{solution1}) suggests otherwise.

Then, we wish to find out what happens inside the collision
region, where the regular Boltzmann equation (\ref{bol}) is
supposed to hold.

The second term on the right side of (\ref{bol}) takes the form
\begin{equation}\label{term2} -\int_{{\bf v}_1}\int_{\Omega_c}
2u f({\bf v})f_1({\bf v}_1) \sigma(\Omega_c)  d\Omega_cd{\bf v}_1.
\end{equation} It is trivial to find that
this integral is computable. However, since (\ref{term2}) is
associated with particles that suffer collisions two times or more
($f$ itself stands for particles produced by collisions), we shall
consider it no more.

After (\ref{term2}) is omitted, (\ref{bol}) becomes, by virtue of
${\partial f}/{\partial t}=0$ and ${\bf F}=0$,
\begin{equation}\label{term1} {\bf v}\cdot \frac{\partial
f}{\partial {\bf r}}=\int_{{\bf v}_1}\int_{\Omega_c} 2u f'({\bf
v}')f'_1({\bf v}'_1) \sigma(\Omega_c)  d\Omega_c d{\bf v}_1.
\end{equation}
Instead of solving (\ref{term1}), the following strategy will be
taken. Under the assumption that $f$ is completely known, say by
experimental means, we try to find out whether or not
(\ref{term1}) is mathematically computable. Seeming peculiar, this
strategy serves us quite well.

The left side can be trivially calculated: at a specified
`position' $({\bf r},{\bf v})$, the side takes a definite and
finite value ($f$ there is regularly differentiable). The
calculation of the right side is nontrivial: to get a definite
value at the same specified `position' $({\bf r},{\bf v})$ is
actually impossible. The basic reason behind the incomputability
is that ${\bf v}'$, ${\bf v}'_1$, ${\bf v}$, ${\bf v}_1$ and
$\Omega_c$ are associated with each other rather solidly. Note
that ${\bf v}$ has been specified and $({\bf v}'+{\bf v}'_1)$
points along the $\pm v_x$ axial in this setup; thus ${\bf v}_1$
has no choice but to lie on the plane determined by the $v_x$
axial and $\bf v$ (under the momentum conservation law).
Similarly, after ${\bf v}$ and ${\bf v}_1$ are specified, there is
no degree of freedom for $\Omega_c$. All these tell us that the
integration domain of (\ref{term1}) is just two-dimensional and
the integral cannot be calculated (See Appendix A, C and Ref.~1
for more analytical discussions).

In fact, there is no great difficulty to configure other types of
setups in which paradoxes concerning Boltzmann's equation show up.

\section{Application of the path-integral approach}
A sharp and inevitable question arises. What are the true
obstacles that keep us from mastering the collective behavior of
classical particles while each of such particles has been
formulated for so long? After a long-time effort to find a way
out, we come to realize that particles in such system involve a
dual nature. Before and after they collide, deterministic
path-equations are obeyed; when they collide with boundaries or
other particles, indeterministic laws have to be invoked ($\sigma$
is of probability in nature). With the motivation to incorporate
path-information into an integral of collision, a new
path-integral approach has been proposed\cite{chen2,chen3}. Here,
we wish to present a simplified version of it.

\includegraphics*[150,545][500,715]{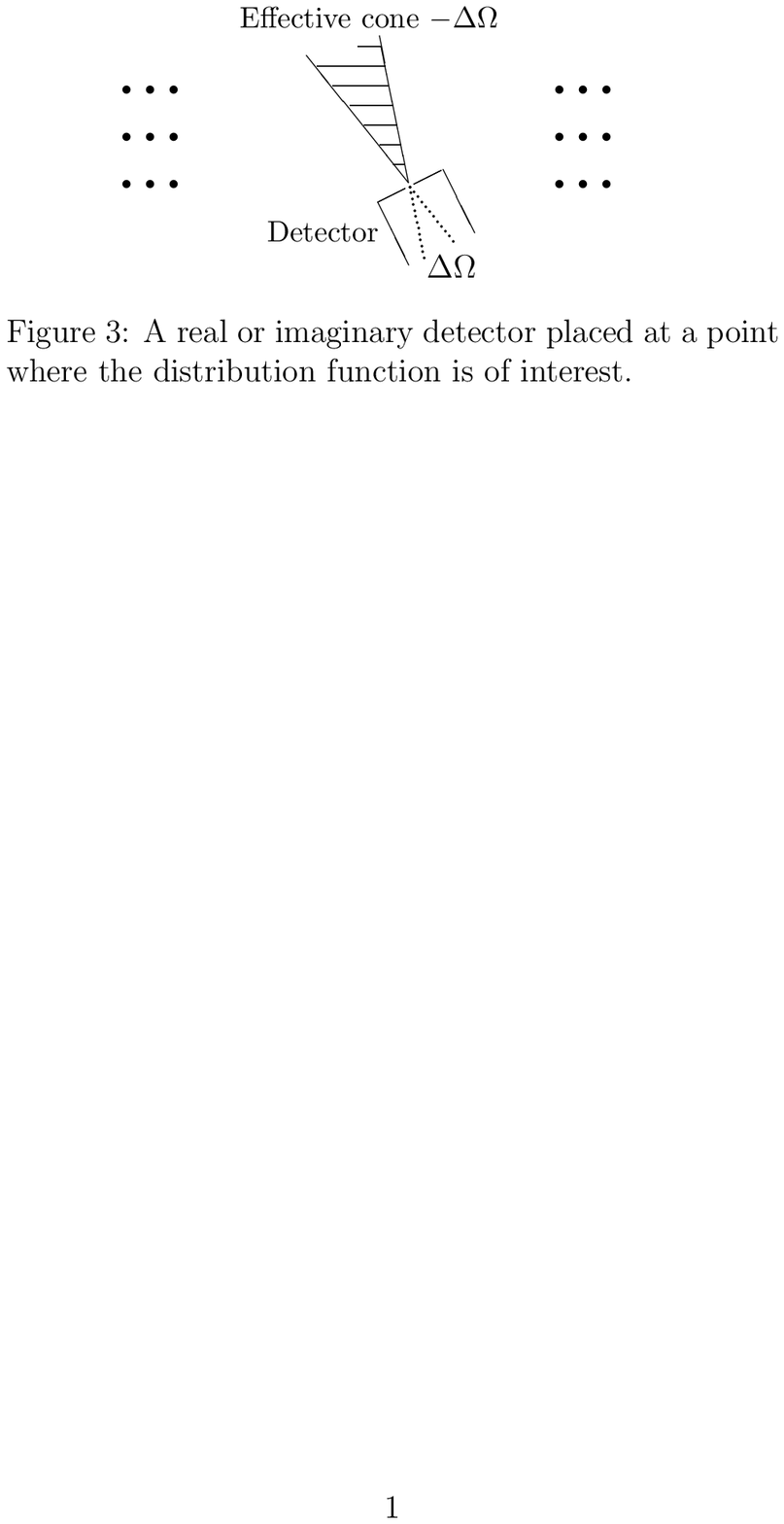}

The primary concern of the proposed theory is focused on what can
be measured in experiments. As in Fig.~3, we consider a particle
detector placed at a position where the distribution function
needs to be determined. There are key points worth mentioning
about the detector.
\begin{enumerate}
\item The detector has a really small opening, whose area is
denoted by $\Delta S_0$.

\item Without specifying the concrete structure of the detector,
it is assumed that every scattered particle within a specific
velocity domain
\begin{equation} \Delta {\bf v}= v^2 \Delta v \Delta
\Omega,\end{equation} will be registered and any others will not.
(The central axis of $\Delta \Omega$ is perpendicular to $\Delta
S_0$.)
 \item While $\Delta S_0$ and $\Delta v$ are considered to be
infinitely small, the domain of $\Delta \Omega$, though also
small, has been assumed to be finite and fixed. As will be seen,
the discrimination against $\Delta \Omega$ is taken almost
entirely from necessity.

\end{enumerate}

If the detector works as described above, do we know the
distribution function at the entry of the detector? The answer is
basically a positive one. If $\Delta N$ is the number that the
detector counts during $\Delta t$, the particle density in the
phase volume element is
\begin{equation} \label{deltaN} f(t,{\bf r}, v,\Delta
\Omega ) \approx \frac {\Delta N} {(\Delta S_0 v \Delta t)(\Delta
v v^2 \Delta \Omega)},
\end{equation}
where ${\bf r}$ is the position of the detector opening. The form
of (\ref{deltaN}) reveals one of the most distinctive features of
this approach: the distribution function is formulated directly.
Apart from other merits, this brings a lot of convenience to the
verification work.

Examining Fig.~3, we see that only the collisions within the
shaded spatial cone $-\Delta\Omega$, called the effective cone
hereafter, can directly contribute to $\Delta N$. Notice that the
effective cone is defined with help of path information.

We shall apply the concepts introduced above to our two setups.

\subsection{Application to setup 1}

Referring to Fig.~1, we wish to determine the distribution
function of scattered particles $f(t,{\bf r},v, \Delta \Omega)$
under the assumption that the detector opening is at $\bf r$ and
there is a piece of the solid surface inside the effective cone
$-\Delta \Omega$.

 One of the essential features related to
particle-to-boundary collisions is that scattered particles will
behave like particles emitted by the surface\cite{kogan}. In terms
of probabilities, the `emission' rate $\rho$ from a surface
element $dS'$, whose position is ${\bf r}'$, can be defined such
that
\begin{equation}\label{surface} d N =\rho (t',{\bf r}',v', \Omega')d t'
d S'd v' d \Omega',\end{equation}  represents the number of the
particles emerging within the solid-angle domain $d\Omega'$ in the
time interval $d t'$ and speed range $dv'$. Generally speaking,
$\rho$ depends on how the particles are `sent' to the boundary,
how many particles get involved and what kind of boundary exists
there. For purposes of this paper, we simply assume that $\rho$
related to each surface element has been known (by experimental
means for instance).

The particles `emitted' from the surface element $d S'$ at $t'$
will enter the detector opening at $t$ and immediately after
getting in they occupy
\begin{equation} |{\bf r}-{\bf r}'|^2 \Delta \Omega_0 v' dt,
\end{equation}
where the solid-angle domain $\Delta \Omega_0$ is defined as
$\Delta \Omega_0= {\Delta S_0}/ {|{\bf r}-{\bf r}'|^2}$; and there
is a time delay between $t'$ and $t$
\begin{equation}\label{timedelay} t=t'+|{\bf r}-{\bf r}'|/v'.
\end{equation}
Thus, according to (\ref{deltaN}), the distribution function at
$\bf r$ takes the form
\begin{equation}\label{fdelta} f(t,{\bf r},v, \Delta \Omega) =
\int_{\Delta S,\Delta v,\Delta \Omega_0}\frac{\rho(t',{\bf r}',v',
\Omega')d t' d S' d v' d \Omega'}{|{\bf r}-{\bf r}'|^2\Delta
\Omega_0 v'  dt\cdot v^2\Delta v\Delta \Omega} ,\end{equation}
where the integration domain $\Delta S$ includes all surface
elements enclosed by the effective cone $-\Delta\Omega$ (with no
blockage assumed). In writing (\ref{fdelta}), we have used the
fact that $\Delta \Omega_0$ is `much smaller' than $\Delta\Omega$,
namely
\begin{equation}\label{dd}\Delta \Omega_0\ll \Delta\Omega,\end{equation}
which has been ensured by the fact that the detector opening $S_0$
is infinitesimal while $\Delta \Omega$ is finite. With help of
(\ref{dd}), every particle which moves from $\Delta S$ and enters
the detector will be considered as one whose velocity is in the
domain $\Delta\Omega$. We finally arrive at
\begin{equation}\label{df} f(t,{\bf r},v,
\Delta \Omega) = \frac{1}{v^3\Delta \Omega}\int_{\Delta
S}\frac{\rho(t',{\bf r}',v, \Omega') d S' }{|{\bf r}-{\bf r}'|^2
}. \end{equation}

It is easy to see that (\ref{df}) can be used to calculate the
behavior of the scattered particles shown in Fig.~1. However, a
much more crucial point is that (\ref{df}) reveals something that
standard approaches have always overlooked: discontinuity almost
everywhere. When two adjacent effective cones pertaining to one
spatial point contain different boundaries, the velocity
distribution there will generally involve discontinuity. In turn,
this means that any finite-size boundary will generate
discontinuity of velocity distribution at every path-reachable
point (collisions along the paths erase such discontinuity partly
and gradually though).

\subsection{Application to setup 2}

Referring to Fig.~2, we now determine the distribution function of
scattered beam 0 particles $f(t,{\bf r}, v, \Delta \Omega)$, where
$\bf r$ is located inside or outside the collision region.

The collisions taking place within $d{\bf r}'$, where ${\bf r}'$
denotes a spatial point inside the effective cone, can be
represented by
\begin{equation} \label{twobeams}
[2ud{\bf r}'][f'(t',{\bf r}',{\bf v}')d{\bf v}'][f'_1(t',{\bf
r}',{\bf v}'_1)d{\bf v}'_1] [\sigma(\Omega_c) \Delta t' ]d
\Omega_c,
\end{equation}
whose derivation can be found in usual textbooks\cite{reif}. If
the particles produced by the collisions expressed by
(\ref{twobeams}) reach $\bf r$ at the time $t$, there must be a
time delay expressed by $ t'=t-{|{\bf r}-{\bf r'}|}/{v}$.

With help of the concept of effective cone and the concept of time
delay, path-information has been fully incorporated. Furthermore,
the two concepts manifest an intriguing relatively new mechanism.
Every physical event inside the effective cone has its `right' to
contribute to the later-time distribution function at $\bf r$,
regardless of its distance to $\bf r$. With this mechanism in
mind, we shall integrate (\ref{twobeams}) over the entire
effective cone. (In our general theory, the same measure is taken
except that each path involves a certain loss probability.)

Since this setup is entirely time-independent, time variables will
be made implicit in all formulas.

\includegraphics*[150,553][500,703]{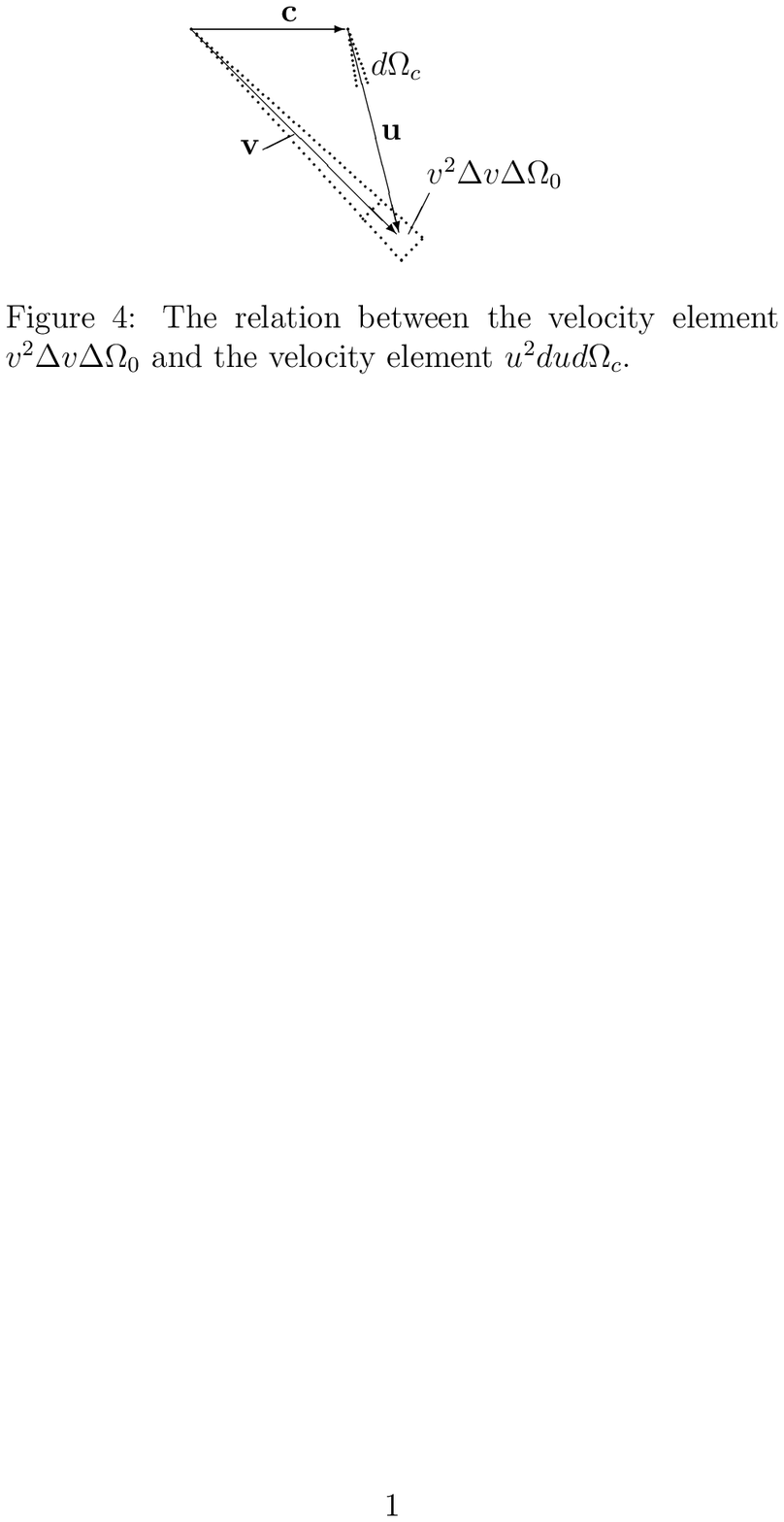}

Inserting (\ref{beam1}) and (\ref{beam2}) into (\ref{twobeams})
and then integrating (\ref{twobeams}), we finally obtain, from
expression (\ref{deltaN}),
\begin{equation} \label{entry} \int_{-\Delta \Omega}d{\bf r}'\int_{\Delta
v,\Delta \Omega_0} d\Omega_c \int_0^\infty dv'_{x}\int_{-\infty}^0
dv'_{1x} \frac{2u\sigma(\Omega_c) n'n'_1  g(v'_{x}) g_1(v'_{1x})}
{(|{\bf r}-{\bf r}'|^2 \Delta \Omega_0 v)\dot (v^2\Delta v \Delta
\Omega )},
\end{equation}
where $\Delta \Omega_0= \Delta S_0/|{\bf r}-{\bf r}'|^2$ and the
integration domain $(\Delta v,\Delta \Omega_0)$ implies that only
the particles in $\Delta v$ and in $\Delta \Omega_0$ can be taken
into account. Again, since $\Delta \Omega_0\ll \Delta \Omega$
holds in most regions inside the effective cone, particles that
can enter the detector will be considered as ones within $\Delta
\Omega$. Based on the notation ${\bf c}'\equiv ({\bf v}'+{\bf
v}'_1)/2$, ${\bf c}\equiv ({\bf v}+{\bf v}_1)/2$, ${\bf u}'\equiv
({\bf v}'-{\bf v}'_1)/2$, ${\bf u}\equiv ({\bf v}-{\bf v}_1)/2$,
$c'\equiv |{\bf c}|$, $c'\equiv |{\bf c}'|$, $u\equiv |{\bf u}|$
and $u'\equiv |{\bf u}'|$, the variable transformation from
$(v'_{x},v'_{1x})$ to $(c',u')$ and finally to $(c,u)$ is made,
and (\ref{entry}) becomes
\begin{equation} \label{entry1} \int_{-\Delta \Omega}d{\bf r}'\int_{\Delta
v,\Delta \Omega_0} d\Omega_c \int_{-\infty}^\infty dc
\int_{|c|}^{\infty} du \frac{2u\sigma(\Omega_c) \|J\|n'n'_1 g(c+u)
g_1(c-u)} {(|{\bf r}-{\bf r}'|^2 \Delta \Omega_0 v)\dot (v^2\Delta
v \Delta \Omega )},
\end{equation}
in which the Jacobian between the variable transformation is
\begin{equation} \| J\|=\left\| {\partial (v'_{1x},v'_{2x})}
/{\partial (c',u')}\right\|=2.
\end{equation}
By investigating the situation in the velocity space shown in
Fig.~4, the following relation can be found out:
\begin{equation} \int_{\Delta v, \Delta \Omega_0}
u^2 d\Omega_c \int_{|c|}^{\infty} du    \cdots \approx v^2\Delta v
\Delta\Omega_0\cdots .
\end{equation}
Therefore, the distribution function at $\bf r$ is
\begin{equation} \label{final} f({\bf r}, v, \Delta
\Omega)=\frac{4n'n'_1}{v \Delta \Omega} \int_{-\Delta \Omega}d{\bf
r}' \int_{-\infty}^\infty dc \frac{\sigma(\Omega_c)g(c+u)
g_1(c-u)}{u|{\bf r}-{\bf r}'|^2},
\end{equation}
where $\Omega_c$ is the solid angle formed by ${\bf u}$ and ${\bf
u}'$ and ${\bf u}$ is defined as ${\bf u}=v ({\bf r}-{\bf r}')
/|{\bf r}-{\bf r}'|-{\bf c}$.

It is very obvious that (\ref{final}) is truly calculable and the
result is a definite number.

Since $\Delta\Omega$ is a finite solid-angle domain, expression
(\ref{final}) is nothing but the distribution function averaged
over $\Delta\Omega$, which is still different from the `true and
exact distribution function' in the standard theory. However, if
we let $\Delta\Omega$ tend to zero, $\Delta\Omega_0\ll
\Delta\Omega$ will no longer hold and the concept of effective
cone will no longer be valid, thus making the whole formalism
collapse. A thorough inspection of this approach compels us to
believe that the true and exact distribution function is beyond
our reach and any attempt to formulate it will ultimately fail.

\section{Conclusions}

The viewpoints and calculations presented in this paper are
waiting to be proved or disproved by real or computational
experiments. If they are right, as earnestly expected by us, many
fundamental and interesting questions arise. For one thing, this
paper hints that differential equations should not be considered
as universally effective apparatuses (as many may assume). The
following observations are worth mentioning.
\begin{itemize}

\item In a differential formalism, we either respect
path-equations, in which ${\bf v}=d{\bf r}/dt$ holds, or respect
partial differential equations, in which $t$, ${\bf r}$ and ${\bf
v}$ are completely independent. In general, we cannot incorporate
path-equations into a partial differential approach. \item In view
of that the newly-formulated distribution function is an averaged
one, the proposed theory is `approximate' in nature and the
time-irreversibility (information loss) has been built in.
 \item
According to the proposed theory, the spatial discontinuity
associated with a boundary will transform itself into the
pervasive velocity discontinuity in the nearby, as well as
distant, region. The role that such discontinuity plays in a
variety of non-equilibrium phenomena has been overlooked.
\end{itemize}

For the time being, these observations, as well as many related
others, are just matters for conjecture. Reference papers can be
found in the regular and e-print literature\cite{chen2,chen3}.

\section*{Acknowledgment}

Communication with Oliver Penrose is gratefully acknowledged. The
author also thanks Hanying Guo, Ke Wu and Keying Guan for helpful
discussion.

\begin{appendix}

\section*{Appendix A: Disproof of Boltzmann's equation}

In this appendix, we shall investigate Boltzmann's equation in a
brief and direct manner (involving no physical assumptions).

According to the philosophy of standard theory\cite{reif}, there
is an expression characterizing how collisions lessen particles
around a specific position $\bf r$ and a specific velocity $\bf v$
\begin{equation}\label{rate1} \lim\limits_{\Delta{t},
\Delta{\bf r},\Delta{\bf v}\rightarrow 0}\frac{ \Delta N_{\rm
out}}{\Delta {t}\Delta {\bf r}\Delta {\bf v}},
\end{equation}
where $\Delta N_{\rm out}$ is the number of the particles that get
out of $\Delta {\bf r}\Delta {\bf v}$ during $\Delta t$ because of
collisions. Likewise, there is an expression characterizing how
collisions produce particles around $\bf r$ and $\bf v$
\begin{equation}\label{rate2} \lim\limits_{\Delta{t},
\Delta{\bf r},\Delta{\bf v}\rightarrow 0}\frac{ \Delta N_{\rm
in}}{\Delta {t}\Delta {\bf r}\Delta {\bf v}},
\end{equation}
where $\Delta N_{\rm in}$ is the number of the particles that get
in $\Delta {\bf r}\Delta {\bf v}$ during $\Delta t$ because of
collisions. Although expressions (\ref{rate1}) and (\ref{rate2})
have served as two of the basic concepts in our understanding to
the statistical world (explicitly or implicitly), the recent
studies of us fall short of justifying them.

In fact, a paradox associated with (\ref{rate1}) can be addressed
directly. To calculate expression (\ref{rate1}), we are supposed
to formulate how many particles initially within $\Delta {\bf
r}\Delta {\bf v}$ collide during $\Delta t$ and thus leave $\Delta
{\bf r}\Delta {\bf v}$. However, if the length scale of $\Delta
{\bf r}$, denoted as $|\Delta {\bf r}|$, be much smaller than
$|{\bf v}\Delta t|$ (in the standard theory there is no constraint
on how $\Delta t$, $\Delta {\bf r}$ and $\Delta {\bf v}$ tend to
zero), we end up with finding that all the considered particles
travel much longer than $|\Delta {\bf r}|$ during $\Delta t$ and
almost all the formulated collisions take place outside $\Delta
{\bf r}$. That is to say, in order for (\ref{rate1}) to hold its
significance, we have no choice but to presume $|\Delta {\bf
r}|\gg |{\bf v}\Delta t|$, which is, unfortunately, in conflict
with Boltzmann's equation itself, whose basic conception is that
$t$, $\bf r$ and $\bf v$ are completely independent of each other.

A similar paradox exists concerning (\ref{rate2}). To calculate
(\ref{rate2}), we need to examine how many particles involve
collisions inside $\Delta {\bf r}$ during $\Delta t$ and acquire
velocities within $\Delta {\bf v}$. However, if $|\Delta {\bf
r}|\ll |{\bf v}\Delta t|$, we find that every examined particle
will, after its collision, escape from $\Delta {\bf r}$ in a time
much shorter than $\Delta t$ and only the particles that collide
at the very end of $\Delta t$ can be possibly regarded as ones
that get in $\Delta {\bf r}\Delta {\bf v}$ during $\Delta t$.

We now turn attention to whether or not (\ref{rate2}) is
computable (with $|\Delta {\bf r}|\gg |{\bf v}\Delta t|$ adopted
for the discussion hereafter). Consider that the gas of interest
consists of many beams and the particles of each beam move at the
same velocity. Following the standard methodology, our task is to
examine how many beam $i$ particles in $\Delta {\bf r}$ will
emerge within the velocity volume element $\Delta {\bf v}$ after
colliding with beam $j$ particles. It is easy to see that although
the scattering velocities ${\bf v}_i$ and ${\bf v}_j$ (of beam $i$
and beam $j$ particles respectively) have all together six
components, they are confined to two degrees of freedom since the
energy-momentum conservation laws serve as four constraints.
Namely, in the velocity space, the ending point of ${\bf v}_i$, as
well as that of ${\bf v}_j$, has to spread over the
energy-momentum shell $S_{ij}$, which is of zero thickness and
defined by the initial velocities ${\bf v}'_i$ and ${\bf v}'_j$
(of beam $i$ and $j$ respectively). Then, a serious problem
related to (\ref{rate2}) is about to surface: the number of the
beam $i$ particles emerging within $\Delta {\bf v}$, labelled as
$(\Delta N_{\rm in})_{ij}$, cannot be proportional to the volume
of $\Delta {\bf v}$, while $(\Delta N_{\rm in})_{ij}$ is trivially
proportional to $\Delta{t}$ and $\Delta{\bf r}$. To this problem,
two possible situations are relevant. The first is one in which
the shell $S_{ij}$ passes through the convergence point of $\Delta
\bf v$ (to which $\Delta\bf v$ shrinks), and therefore
\begin{equation}\label{rateij1}
\lim\limits_{\Delta{t}, \Delta{\bf r},\Delta{\bf v}\rightarrow
0}\frac{ (\Delta N_{\rm in})_{ij}}{\Delta {t}\Delta {\bf r}\Delta
{\bf v}} \propto \lim\limits_{a\rightarrow 0}\frac{\sigma a^2}
{a^3}\rightarrow \infty,
\end{equation}
where $\Delta {\bf v}$ has been assumed to be cube-shaped with
side length of $a$ and $\sigma$ is the local surface density of
beam $i$ particles on $S_{ij}$. The second is one in which
$S_{ij}$ does not pass through the convergence point of $\Delta
\bf v$, and therefore no beam $i$ particle will emerge within
$\Delta {\bf v}$ as $\Delta {\bf v}\rightarrow 0$, i.e.
\begin{equation}\label{rateij2} \lim\limits_{\Delta{t},
\Delta{\bf r},\Delta{\bf v}\rightarrow 0}\frac{ (\Delta N_{\rm
in})_{ij}}{\Delta {t}\Delta {\bf r}\Delta {\bf v}} \rightarrow 0.
\end{equation}
Expressions (\ref{rateij1}) and (\ref{rateij2}) show that the rate
(\ref{rate2}) is not associated with a definite value even after
$|\Delta {\bf r}|\gg |{\bf v}\Delta t|$ is adopted. (The
manipulation of the standard theory makes the problem less
visible, but not absent. See Sect.~3 and Appendix C for more.)

\section*{Appendix B: Two types of cross sections}
To understand Boltzmann's equation, it is of considerable
necessity to know the differences between the cross section in the
center-of-mass frame and the cross section in the laboratory
frame.

First of all, we examine the collisions between a particle beam
(of type 0) and a single particle (of type 1) in the
center-of-mass frame. Let $\Omega_c$ denote the solid angle formed
by the velocity of a scattered type 0 particle with respect to its
initial velocity. Then, the cross section $\sigma(\Omega_c)$ is
defined such that
\begin{equation}\label{sigma} dN=\sigma(\Omega_c)d\Omega_c
\,\,[{\rm or}\,\,dS_c=\sigma(\Omega_c) d\Omega_c],\end{equation}
represents the number of the scattered type 0 particles emerging
within the solid-angle domain $d\Omega_c$ per unit time and per
unit relative flux. Since the particles emerging within
$d\Omega_c$ actually come from the incident area $dS_c$, the cross
section has the dimension and units of area. As textbooks have
clearly elaborated, $\sigma(\Omega_c)$ can be applied normally and
easily\cite{landau,reif}.

Now, we examine the cross section in the laboratory frame. In a
seemingly identical way, consider the collisions between a
particle beam (of type 0) with velocity ${\bf v}'$ and a single
particle (of type 1) with velocity ${\bf v}'_1$. According to
textbooks of statistical mechanics, there is a cross section
$\hat\sigma$ such that\cite{reif}
\begin{equation} \label{sigma1}
dN= \hat\sigma({\bf v}',{\bf v}'_1 \rightarrow {\bf v}, {\bf v}_1
) d{\bf v} d{\bf v}_1
\end{equation}
represents the number of the scattered type 0 particles emerging
between ${\bf v}$ and ${\bf v}+d {\bf v}$, while the type 1
particle emerges between ${\bf v}_1$ and ${\bf v}_1+d {\bf v}_1$,
per unit time and per unit relative flux. At first glance, like
$\sigma(\Omega_c)$, $\hat\sigma$ in (\ref{sigma1}) can be applied
freely. Our studies, however, reveal that $\hat\sigma$ possesses
truly misleading features.

As well known, the energy-momentum conservation laws are obeyed in
a collision. Assuming all colliding particles to have the same
mass (simply for simplicity) and adopting the notation introduced
in Sect.~4, we have
\begin{equation}\label{ene-mom} {\bf c}={\bf c}'\equiv {\bf
c}_0\quad {\rm and }\quad u=u'\equiv u_0, \end{equation} where
${\bf c}_0$ is the velocity of the center-of-mass and $u_0$ the
speed of particles relative to the center-of-mass. For any
specific pair $({\bf v}',{\bf v}'_1)$, ${\bf c}_0$ stands for
three constants and $u_0$ for one, and therefore the degrees of
freedom for $({\bf v},{\bf v}_1)$ are not six but two, as shown in
Fig.~5. Namely, in the velocity space, the velocities ${\bf v}$
and ${\bf v}_1$ have no choice but to fall on the spherical shell
of radius $u_0$, called the energy-momentum shell in this paper.
With help of this concept, we concluded in Ref.~1 that scattered
particles should be examined with reference to an area element on
the energy-momentum shell rather than with reference to a velocity
volume element.

\includegraphics*[150,525][500,720]{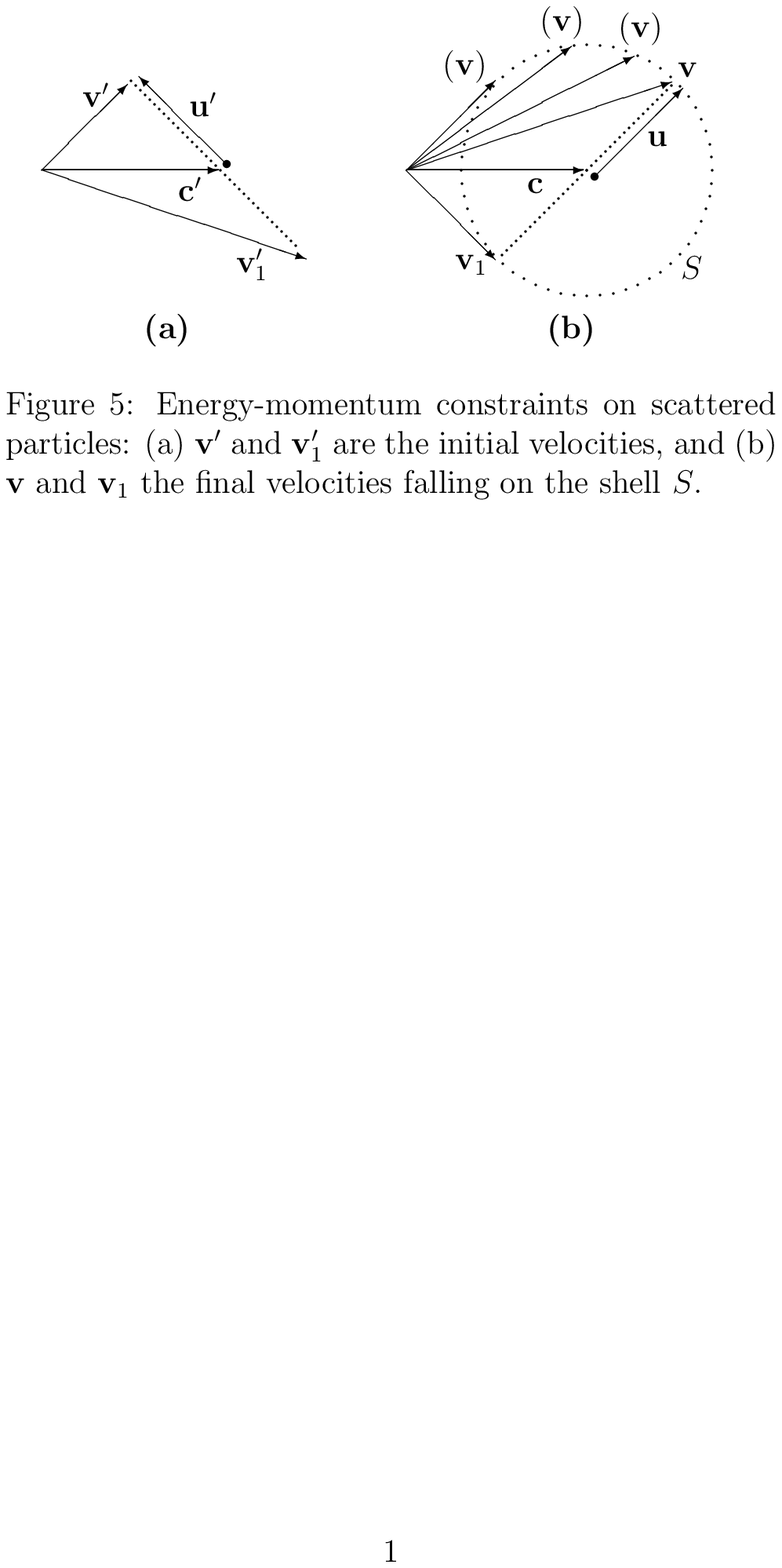}

However, some textbooks state that the cross section $\hat\sigma$
makes sense in terms of $\delta$-functions\cite{kubo}. To see
whether it is possible and what is meant by it, we rewrite
(\ref{sigma1}) as
\begin{equation} \label{sigma2} dN=\hat\sigma d{\bf v} d{\bf v}_1
=\hat\sigma\|J\|d{\bf c} d{\bf u}=\hat\sigma\|J\| d{\bf c} u^2 du
d\Omega_c,
\end{equation}
where $\|J \|=\|\partial ({\bf v},{\bf v}_1)/\partial ({\bf
c},{\bf u})\|=8$ is the Jacobian between the two systems. The
comparison between (\ref{sigma2}) and (\ref{sigma}) yields
\begin{equation} \label{sigmaf}
\hat\sigma ({\bf v}',{\bf v}'_1\rightarrow {\bf v},{\bf
v}_1)=\frac 1{u^2\|J \|} \sigma(\Omega_c) \delta^3({\bf c}-{\bf
c_0})\delta(u-u_0),
\end{equation}
where $\delta^3$ is the symbol of three-dimensional
$\delta$-function. Obviously, the energy-momentum conservation
laws have been included in (\ref{sigmaf}).

Can the cross section $\hat\sigma$ so defined offer extra
advantages? Or, on second thoughts, can it still cause ill
treatments in certain contexts? To answer these questions, we look
at Fig.~6, in which two small domains for scattered particles are
specified as $\Delta {\bf v}$ and $\Delta {\bf v}_1$ (initial
beams with ${\bf v}'$ and ${\bf v}'_1$ have not been depicted).
Under the assumptions that $\Delta {\bf v}$ and $\Delta {\bf v}_1$
are symmetric with respect to the center-of-mass and $\Delta {\bf
v}$ encloses a small part of the energy-momentum shell, whose area
is denoted by $\Delta S$, we obtain, by virtue of (\ref{sigma2})
and (\ref{sigmaf}),
\begin{equation} \label{sec2.6}\int_{\Delta{\bf v},\Delta{\bf v}_1}
\cdots \hat \sigma d{\bf v} d{\bf v}_1= \int_{\Delta \Omega_c}
\cdots \sigma(\Omega_c)d\Omega_c=\int_{\Delta S} \cdots
\sigma(\Omega_c)dS/u^2 ,
\end{equation} where $\Delta\Omega_c$ is
associated with $\Delta{\bf v}$, via $\Delta\Omega_c=\Delta
S/u^2$. Equation (\ref{sec2.6}) illustrates that $\hat \sigma$ is
essentially defined on the energy-momentum shell $S$ and by the
cross section $\sigma(\Omega_c)$.

\includegraphics*[150,530][500,725]{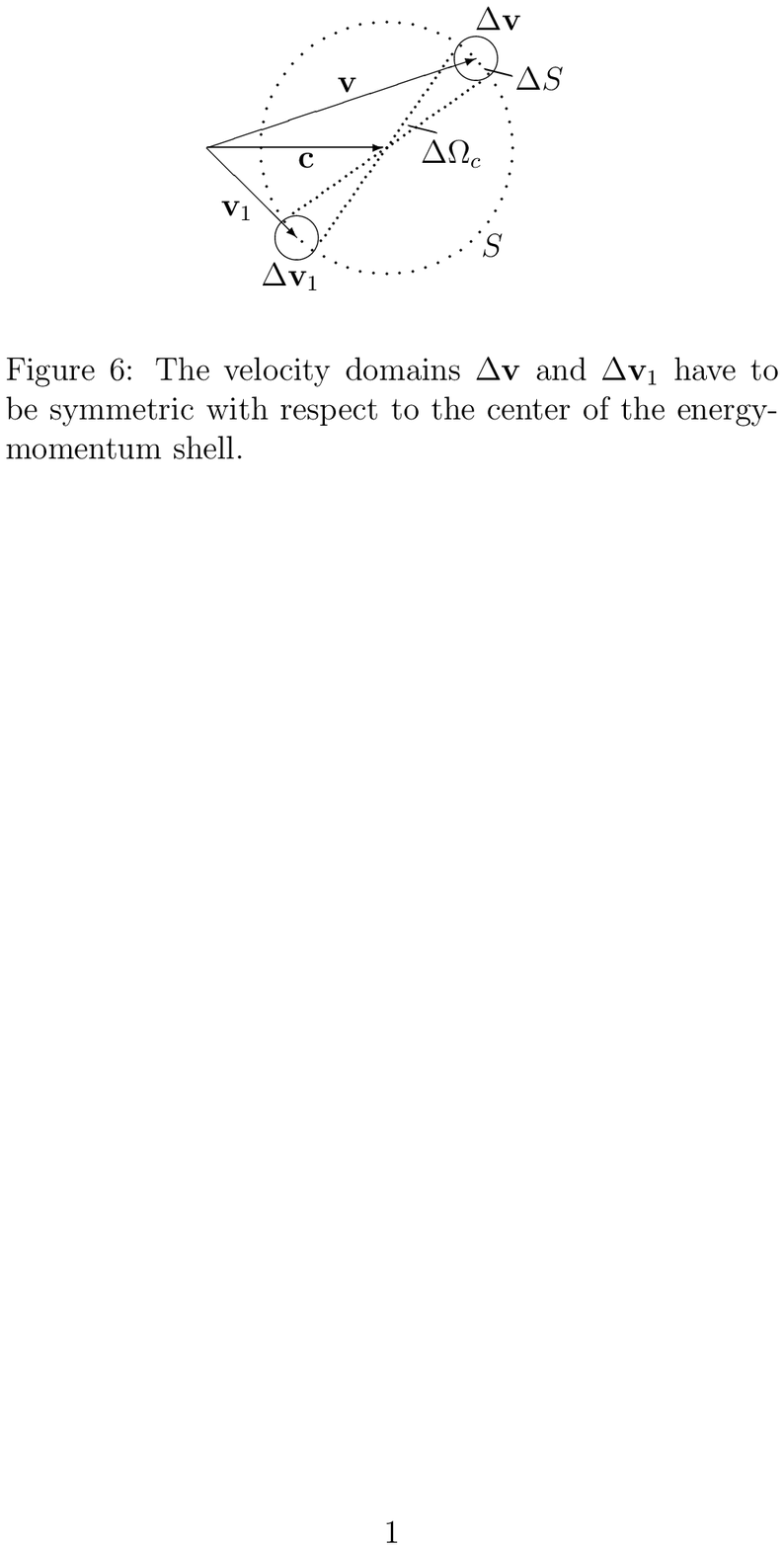}

If there exists no explicit constraint on  ${\bf v}$ and ${\bf
v}_1$ (other than the energy-momentum conservation laws),
expression (\ref{sec2.6}) becomes
 \begin{equation} \label{sec2.7}\int_{{\bf v},{\bf v}_1}\cdots
 \hat \sigma d{\bf v} d{\bf v}_1
=\int_{(4\pi)} \cdots \sigma(\Omega_c)d\Omega_c, \end{equation}
where the subindex $(4\pi)$ of the last integral means that the
integral is over the entire domain of $\Omega_c$.

If $\Delta {\bf v}$ encloses a small part of the energy-momentum
shell $\Delta S=u^2\Delta \Omega_c$ while ${\bf v}_1$ is subject
to no explicit constraint, then the expression
\begin{equation}\label{sec2.8}\int_{\Delta{\bf v},{\bf v}_1}\cdots \hat
\sigma d{\bf v} d{\bf v}_1   \end{equation} seems to differ from
(\ref{sec2.6}) in an obvious way. However, the difference is a
misleading one. Due to the existence of energy-momentum
conservation laws, the confinements of ${\bf v}$ and ${\bf v}_1$
are intrinsically connected, and therefore (\ref{sec2.8}) yields
exactly the same result as (\ref{sec2.6}) does.

In Appendix C, it will be unveiled that, associated with
mistreating (\ref{sec2.8}), the standard formalism involves a
mathematical error.

\section*{Appendix C: An error in the textbook treatment}

According to textbooks, the regular Boltzmann equation is based
on\cite{reif}
\begin{equation}\label{bolz2} \frac{\partial f}{\partial t}+
{\bf v}\cdot\frac{\partial f}{\partial {\bf r}}+ \frac{{\bf
F}}m\cdot \frac{\partial f} {\partial {\bf v}}=\lim_{\Delta
t\Delta {\bf r}\Delta{\bf v} \rightarrow 0}\frac{(\Delta N)_{\rm
in}-(\Delta N)_{\rm out}}{\Delta t\Delta {\bf r}\Delta{\bf v}},
\end{equation} where $(\Delta N)_{\rm in}$ and $(\Delta N)_{\rm out}$
stand for the particles that enter and leave $\Delta {\bf
r}\Delta{\bf v}$ during $\Delta t$ due to collisions. In this
appendix, we ignore the paradoxes unveiled in Appendix A and
concern ourselves solely with how textbooks treat (\ref{bolz2}) in
the mathematical sense.

The textbook methodology of formulating $(\Delta N)_{\rm out}$ is
that, for a definite $\Delta {\bf r}$, the particles leaving
$\Delta{\bf v}$ during $\Delta t$ due to collisions are identified
as $(\Delta N)_{\rm out}$. On this understanding, $f_1({\bf v}_1)
d{\bf v}_1\Delta {\bf r}$ represents the particles knocking some
particles out of the beam $f({\bf v}) d{\bf v}$, and we get
\begin{equation}\label{in11} (\Delta N)_{\rm out}=
\int_{\Delta {\bf v},{\bf v}_1, {\bf v}',{\bf v}'_1} 2u\Delta {\bf
r}[ f({\bf v}) d{\bf v}] [f_1({\bf v}_1)d{\bf v}_1 ][\hat\sigma
\Delta t] d{\bf v}'d{\bf v}'_1, \end{equation} where
$\hat\sigma=\hat\sigma({\bf v},{\bf v}_1\rightarrow {\bf v}',{\bf
v}'_1) $ is the cross section in the laboratory frame. With help
of (\ref{sec2.7}), we get
\begin{equation}\label{a3}
\lim_{\Delta t\Delta {\bf r}\Delta{\bf v} \rightarrow 0}
\frac{(\Delta N)_{\rm out}}{\Delta t\Delta {\bf r}\Delta{\bf v}} =
\int_{{\bf v}_1, \Omega_c(4\pi) } 2u \sigma(\Omega_c) f({\bf v})
f_1({\bf v}_1) d{\bf v}_1 d\Omega_c,
\end{equation} where ${\bf v}_1$ is unlimited and the domain of
$\Omega_c$ is ($4\pi$). So far, every thing is exactly the same as
that in the textbook treatment.

A similar methodology is employed to formulate $(\Delta N)_{\rm
in}$. $f'({\bf v}')d{\bf v}'$ and $f'_1({\bf v}'_1)d{\bf v}'_1$
are identified as two colliding beams, and we have
\begin{equation}\label{in22} (\Delta
N)_{\rm in}=\int_{{\bf v}',{\bf v}'_1,\Delta {\bf v}, {\bf v}_1}
2u\Delta {\bf r} [f'({\bf v}') d{\bf v}'] [f'_1({\bf v}'_1)d{\bf
v}'_1 ][\hat\sigma \Delta t] d{\bf v}d{\bf v}_1,
\end{equation}
where $\hat\sigma=\hat\sigma({\bf v}',{\bf v}'_1 \rightarrow {\bf
v}, {\bf v}_1 )$ and $\Delta {\bf v}$ is used to remind us that
only the particles emerging within $\Delta {\bf v}$ will be taken
into account. Although the symmetry between (\ref{in11}) and
(\ref{in22}) appears obvious and has played a vital role in
deriving the Boltzmann equation, we are now convinced that it is
an illusive one. Due to the existence of the energy-momentum
conservation laws, when ${\bf v}$ in (\ref{in22}) is limited to an
infinitesimal volume $\Delta{\bf v}$, ${\bf v}_1$ must be limited
to the symmetric $\Delta {\bf v}_1$, while ${\bf v}_1$ in
(\ref{a3}) is unlimited. That is to say, (\ref{in22}) should be
replaced by
\begin{equation}\label{in221} (\Delta N)_{\rm in}=
\int_{{\bf v}',{\bf v}'_1,\Delta {\bf v}, \Delta{\bf v}_1}
[2u\Delta {\bf r}] [f'({\bf v}') d{\bf v}'] [f'_1({\bf v}'_1)d{\bf
v}'_1 ][\hat\sigma \Delta t] d{\bf v}d{\bf v}_1.
\end{equation}
So, the aforementioned symmetry disappears and we arrive at
\begin{equation}\label{a4}
\lim_{\Delta t\Delta {\bf r}\Delta{\bf v} \rightarrow 0}
\frac{(\Delta N)_{\rm in}}{\Delta t\Delta {\bf r}\Delta{\bf v}}
\not= \int_{{\bf v}_1, \Omega_c(4\pi) } 2u \sigma(\Omega_c)
f'({\bf v}') f'_1({\bf v}'_1) d{\bf v}_1 d\Omega_c.
\end{equation}

As Appendix A has shown, a serious evaluation of the left side of
(\ref{a4}) actually leads us to nowhere.
\end{appendix}

\end{document}